# Memory effect under pressure in low density amorphous silicon

Nandini Garg, K.K. Pandey, K.V. Shanavas, C. A. Betty[†] and Surinder M Sharma
High Pressure and Synchrotron Radiation Physics Division, [†]Chemistry Division,
Bhabha Atomic Research Centre, Mumbai 400085, India

***Abstract***

Our investigations on porous Si show that on increase of pressure it undergoes crystalline phase transitions instead of pressure induced amorphization - claimed earlier, and the amorphous phase appears only on release of pressure. This amorphous phase, when subjected to higher pressures, transforms *reversibly* to a higher coordinated primitive hexagonal phase showing a kind of memory effect which may be the only example of its kind in the elemental solids. First principles calculations and thermodynamic arguments help understand these observations.

[*]Corresponding author : Fax: +91-22-25505151

Tel.: +91-22-25595476

Email address: nandini@barc.gov.in (Nandini Garg)

In the crystalline form, depending upon the size of the crystallites, Si undergoes structural modifications to higher coordinated denser forms under pressure.[1-8] For example cubic diamond phase of nano-crystalline Si (~ 50 nm) transforms directly to the primitive hexagonal (**ph**) form, by-passing several structures observed in the bulk.[5] For still smaller sized crystallites, such as in porous Si (~ 4-5 nm), the initial cubic phase has been claimed to transform to a high density amorphous (HDA) form which on release of pressure was shown to convert to a low density amorphous (LDA) structure.[9] Some of these observations have also been supported by theoretical calculations.[10-19] Experiments on bulk amorphous silicon have shown that it transforms reversibly to metallic HDA at high pressures .[20]

In this letter we report two new observations: i) cubic nano porous Si (π-Si) undergoes a crystalline phase transition to primitive hexagonal [1] (**ph**) phase under high pressure and the amorphous phase arises only when the pressure is reduced from the stability field of **ph** phase,( Hereafter referred to as nano – amorphous ) ii) this nano- amorphous phase transforms reversibly to the **ph** phase under pressure. We also present theoretical calculations based on density functional theory and simple nucleation growth model to explain the experimental observations. Moreover, this observed amorphous-crystalline reversible inter-conversion is also observed in bulk amorphous Si, though the pressure release needs to be fast unlike in the case of nano – amorphous Si.

The experimental investigations presented here were carried out on π-Si as well as bulk amorphous Si. π-Si was prepared through electrolytic etching.[21] The particle size, determined from x-ray diffraction as well as Raman results (quantum confinement model), was ~ 4-5 nm.[22] X-ray structure factor (S(**q**)) of amorphous Si at ambient conditions was found to be consistent with tetrahedral coordination. 99.9% pure amorphous silicon samples were obtained from BHEL (India) and were characterized by FTIR and Raman spectroscopy.

For high pressure experiments the sample was loaded in a ~150 μm hole, drilled in a pre-indented tungsten gasket of a diamond-anvil cell (DAC). X-ray diffraction experiments on π-Si, were carried out at BL10XU of Spring8 synchrotron, using a x-ray beam of ~ 30 μm (diameter) and λ= 0.3085 Å. 4:1 methanol-ethanol mixture was used as a pressure transmitting medium and pressure was estimated using ruby fluorescence method. For amorphous Si, in addition to the measurements at SPring8, some of the experiments were also performed at XRD1 beamline of Elettra synchrotron. For these experiments a few particles of gold were also loaded in the DAC to determine the pressure in the cell. Raman measurements on both the samples were carried out in our laboratory using micro-Raman set-up. The Raman modes were excited with the solid state pumped laser of wavelength 532 nm.

On increase of pressure, Raman results, shown in Fig. 1, confirm vanishing of Raman active peak of cubic phase of π-Si beyond 18 GPa. For example, at 23 GPa we get a flat background. This result is very similar to the results published earlier,[9] which had been interpreted as pressure induced amorphization, rationalized in terms of pressure induced pseudo-melting due to negative Clapeyron slope .[2,23,24] However, our x-ray diffraction results, given in Fig. 2, show that π-Si does not become amorphous at least upto ~ 36 GPa. Instead it transforms to 8-coordinated **ph** phase at ~ 20 GPa. At this pressure the most intense [100] and [101] diffraction peaks of the **ph** phase are clearly visible. These observations are similar to those of Tolbert *et al*.[5] The coexistence of the diamond as well as the **ph** phase of π-Si may be attributed to the inhomogeneous distribution of local stresses in the sample, aided by pores etc. Non-observability of Raman mode beyond 18 GPa is also consistent with this (as **ph** phase does not have any optical phonon) and does not imply an amorphous phase. A small difference in the pressure of transformation from that of Tolbert et al[5] may be understandable in terms of different topology and size of nano-particles. We may also note that in MD simulations for bulk Si, cubic diamond phase has been shown to transform to the **ph** phase on abrupt pressure increase[25]. It is also known that β–tin phase becomes inaccessible when Si is compressed at low temperatures ( < 100 K).[26] These facts suggest the existence of a low barrier path between cubic diamond and ph phase.

Our Raman results (Fig. 1) show that on release of pressure the primitive hexagonal phase obtained from π–Si transforms to a low density amorphous (LDA) phase through a high density amorphous phase (HDA), consistent with the earlier studies.[5] The broad band centered at 392 $cm^{-1}$ at ~ 15 GPa , can be assigned to the HDA phase. Emergence of HDA phase in the vicinity of the stability region of the β-tin phase implies that the amorphous phase in the present case is due to kinetic preference.[27] On further release of pressure the low density amorphous phase was identifiable at ~ 4.5 GPa, characterized by a broad hump of tetrahedral Si-Si stretching vibrations at 484 $cm^{-1}$. When this LDA phase was re-pressurized, it transformed back to the **ph** phase at ~ 18 GPa, implying reversible amorphous to crystalline phase transformation under pressure. From Fig. 1 it is evident that even in the second pressure cycle the LDA-**ph** and reverse transformation proceeds through a HDA phase. In fact when pressure was released in the second pressure cycle a coexistence of the LDA and HDA phases was observed at ~ 7.5 GPa.

Our x-ray diffraction (XRD) results are also in agreement with the Raman measurements. On release of pressure the XRD pattern shows that a small fraction of the amorphous phase was observable even at 12 GPa.[28] On re-pressurizing the amorphous phase (Fig. 2) the diffraction peaks of **ph** phase could be observed at ~ 18 GPa. On full release of pressure the sample stays fully amorphous (Fig. 2). Hence both XRD and Raman results confirm the pressure induced reversibility between the amorphous and **ph** phases in π–Si. Here we would like to mention that no elemental solid is known to display this kind of memory effect between a crystalline and an amorphous phase.

On subjecting the bulk amorphous Si to high pressure, it also transforms to the primitive hexagonal phase, preceded by LDA-HDA transformation, similar to nano – amorphous Si. However, in this case the crystallization takes place at a lower pressure of ~ 14 GPa. Interestingly, bulk amorphous phase exhibits significant deviation during decompression process: if the pressure is reduced slowly, it follows same sequence of transformations as the bulk crystalline Si i.e. appearance of ***β*-Sn** phase and subsequent **R-8** phase on full release. If decompressed fast enough (< ~ 60 secs) we get the amorphous (LDA) phase. Re-pressurizing this LDA again takes it back to the **ph** phase following same transformation sequence i.e., LDA-HDA-**ph**. To check whether the observed reversible amorphous-**ph** transition is an inherent property of amorphous Si or it is specific to the LDA phase obtained from **ph** phase, we prepared another amorphous (LDA) Si sample from bulk crystalline Si following pressure-temperature cycle suggested by Imai et al[29] (with ***β*-Sn** phase as a precursor to LDA-Si). Our experiments confirm the same reversible nature of amorphous-ph-amorphous phase transition.

The preferred crystallization of **ph** from HDA can be understood in terms of thermodynamic arguments used earlier by others[30] to understand pressure induced crystallization of glass. For quantitative estimates of pressures of crystallization of competing phases, we used the interfacial energy, equation of state of HDA[13] and of crystalline phases[31] obtained from the DFT calculations. Using these, it can be shown[32] that HDA should transform to ***β*-Sn** at ~ 12 GPa and to the **ph** phase at ~18 GPa. However, the first principle calculations of ref. 13 show that LDA-HDA transformation takes place above 15 GPa. Therefore, HDA transforms preferentially to **ph** phase as the crystallization to the ***β*-Sn** phase is prohibited due to the fact that it would take place at a pressure which is lower than that of HDA formation.[33]

To gain further insights, particularly in context of kinetics independent transformation of **ph** phase ( obtained from π-Si) to LDA, several first principles calculations were also carried out on Si nano clusters. DFT calculations were used to optimize ionic positions of small silicon clusters of approximately ~110 atoms and diameter ~1.4 nm. Structural relaxations were performed within Generalized Gradient Approximation (GGA)[34] using Projector Augmented Wave (PAW)[35] method as implemented in Vienna *ab-initio* simulation package (VASP)[36] starting from spherical clusters. An energy cutoff of 400 eV with gamma point sampling was used to find the lowest energy configurations of clusters in a 20 Å supercell. The amorphous structure was generated by heating cubic silicon cluster to 1500 K and quenching it. Molecular Dynamics simulation was then carried out for 0.3 ps at 300 K to equilibrate the cluster and was then subsequently optimized at 0 K. The resultant cluster had almost the same volume as the diamond structure.

The structures of the different phases of silicon, obtained after structural optimization, are shown in Fig. 3. Our results, indicate that for these small clusters, amorphous phase has lower energy than even the diamond structure (difference in Energy/atom = 67.9 meV). We found that the β-tin structure does not stabilize in small clusters whereas diamond and simple hexagonal structures are retained under structural optimization. This could explain the non-observability of the β-tin phase in π-Si. A similar size dependent crystallization has been observed recently in silver.[37]

In conclusion, this is the first observation of a pressure induced reversible amorphous-crystalline transformation in an elemental solid and should encourage further experimental as well as theoretical studies. π-Si does not amorphize on compression and instead it undergoes a crystalline transformation and the amorphous phase arises only on decompression. Moreover, our studies show that irrespective of the method of preparation, the amorphous Si always transforms to the primitive hexagonal phase under compression.

References

1. H. Olijynk, S.K. Sikka and W. B. Holzaphel, *Phys. Lett*., **103A**, 137, (1984)
2. McMillan P. F., Wilson M., Wilding M. C., Daisenberger D., Mezouar M. and Greaves G. N., *J. Phys.: Cond. Mat*. **19** , 415101 (2007)
3. Jamieson J.C., *Science*, **139**, 762, (1963)
4. Wentorf R. H. and Kasper J. H., *Science*, **139**, 338 ( 1963)
5. Tolbert S. H., Herhold A. B., Brus L. E., A.P. Alivisatos, *Phys. Rev. Lett.,* **76**, 4384, (1996)
6. Tolbert S. H., A.P. Alivisatos, *Annu. Rev. Phys. Chem.,* **46**, 595, (1995)
7. Poswal H. K., Garg Nandini, Sharma Surinder M., Busetto E., Sikka S. K., Gundiah G., Deepak F. L., and Rao C. N. R., *J. of Nanoscie. and Nanotech.*, **X**, 1, (2005)
8. V. Domnich and Y. Gogotsi, *Rev. Adv. Mat. Sci*. **3**, 1, (2002)
9. Deb S.K., Wilding M. C., Somayazulu M. & McMillan P. F. *Nature* **414,** 528–530 (2001).
10. Behler J., Martonˇa´k R., Donadio D., and Parrinello M, *Phys. Rev. Lett.* **100**, 185501 ,(2008)
11. Durandurdu M. and Drabold D. A., *Phys. Rev. B*, **66**, 205204 (2004)
12. Durandurdu M. and Drabold D. A., *Phys. Rev. B*, **67**, 212101 (2003)
13. Duranduru, M. & Drabold, D. A. *Phys. Rev. B* **64,** 014101 (2001)
14. Daisenberger D., Wilson M., McMillan P. F., Cabrera R. Q., Wilding M. C., Machon D., *Phys. Rev. B* **75**, 224118 ,(2007)
15. R. Martoˇnák, C. Molteni, and M. Parrinello, *Phys. Rev. Lett.* **84**, 682 ,(2000)
16. Nagy K. G., Schmitt M., Pavone P., Strauch D., *Comp. Mat. Sci*., **22**, 49, (2001)
17. Tetsuya Morishita, *Phys. Rev. Lett.* **93**, 055503 ,(2004)
18. Brad D. Malone, Jay D. Sau, and Marvin L. Cohen, *Phys. Rev. B,* **78**, 035210 (2008)
19. P. Focher, G. L. Chiarotti , M. Bernasconi, E. Tosatti and M. Parrinello, *Europhys. Lett.,* **26,** 345 (1994)
20. McMillan P. F., Wilson M., Daisenberger D., Machon D., *Nature mat.*, **4**, 680, (2005), and references therein
21. Xie, Y. H. et al.. *Phys. Rev. B* **49**, 5386 (1994).
22. Mavi H.S., Rasheed B. G., Shukla A. K., Abbi S. C., and Jain K. P., *J. Phys. D. Appl. Phys.*, **34**, 292 (2001)
23. Mishima O., L. D. Calvert & E. Whalley, *Nature* **310**, 393-395 (1984)
24. Hemley R.J., Jephcoat A. P., Mao H. K., Ming L. C. & Manghnani M. H., *Nature (Lond.)***,** **334**, 52 (1988)
25. Focher P., Chiarotti G. L., Bernasconi M., Tosatti E. and Parrinello M., *EuroPhys. Lett*., **26**, 345 (1994)
26. Imai M., Mitamura T., Yaoita K., Tsuji K., *High Press., Res.,* **15**, 167, (1996)
27. Sharma Surinder M. and Sikka S. K., *Progress in Mat. Sci*., **40**, 1, (1996)
28. To ensure that the sample had not moved out of the region exposed to x-rays, the diamond cell (at ~ 11 GPa) was moved such that x-rays grazed the edge of the gasket hole. The appearance of the gasket peaks is a clear indication that x-rays bathe the sample and hence it is indeed amorphous.
29. Motoharu Imai, Takeshi Mitamura, Kenichi Yaoita, Kazuhiko Tsuji, *High Press. Res.*, **15**, 167 (1996)

30. F Ye and K. Lu, *Phys. Rev. B*, **60**, 7018 (1999)
31. Shanavas et al, to be published
32. K.K. Pandey et al, The details of these calculations, which deduce the crystallization in terms of critical size of nuclei under pressure, will be presented elsewhere.
33. For consistency all the parameters of the model have been taken from the first principles calculations (i.e., references 13, 31)
34. FU Huaxiang, XIE Xide, *Chin. Phys. Lett*., **12**, 245 (1995)
35. Blöchl PE: *Phys. Rev. B*, **50**,17953 (1994)
36. Kresse G: Phys. Rev. B, **54**, 11169 (1996)
37. Jixiang Fang, Hongjun You, Peng Kong, Bingjun Ding, and Xiaoping Songa_, *Appl. Phys. Lett.,* **92**, 143111 (2008)

**Figure 1**

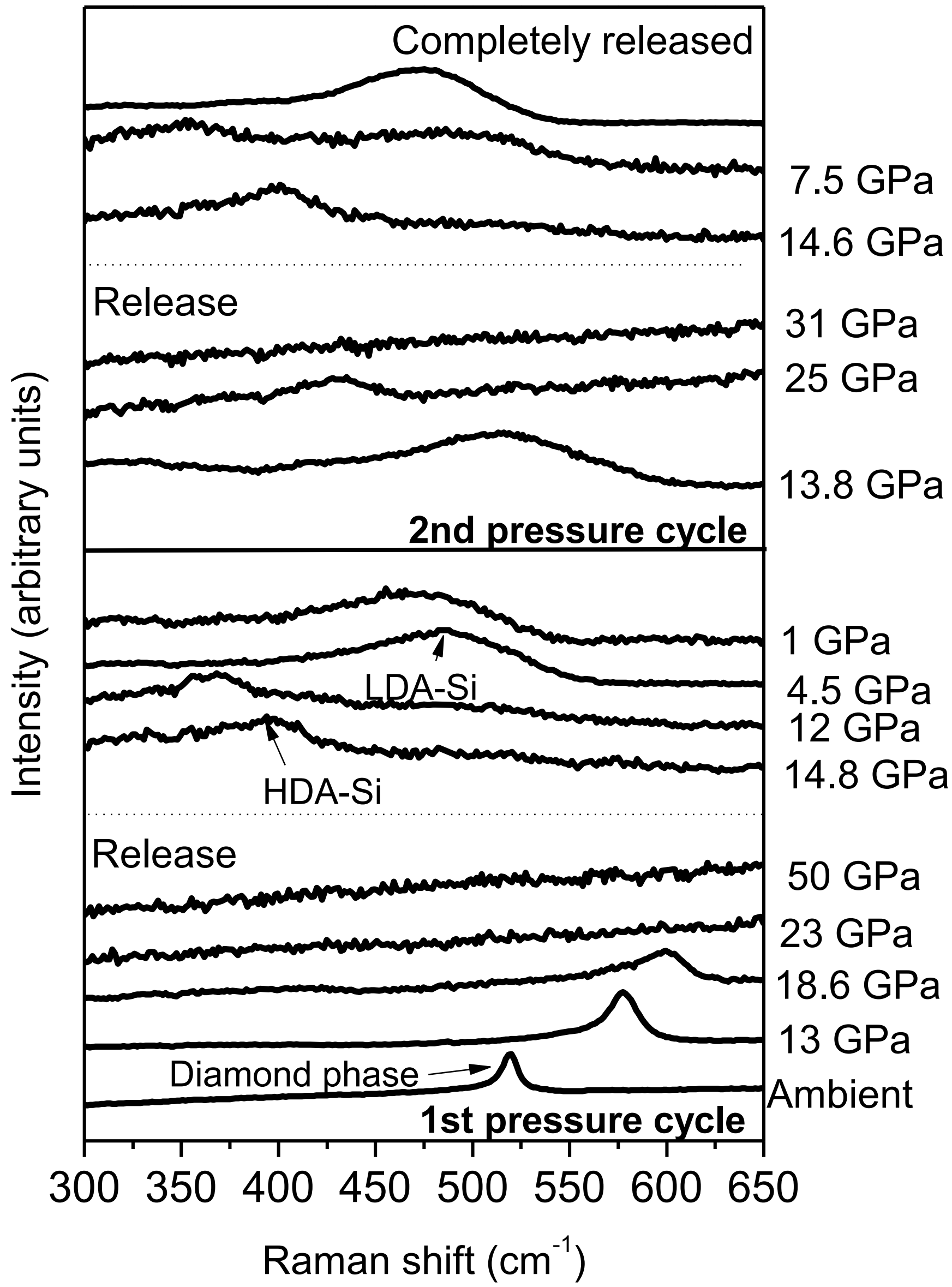

Fig. 1 The Raman spectra of π-Si at different pressures. At ambient pressure the Raman peak of the diamond phase can be seen at 519.54 $cm^{-1}$.

# Figure 2

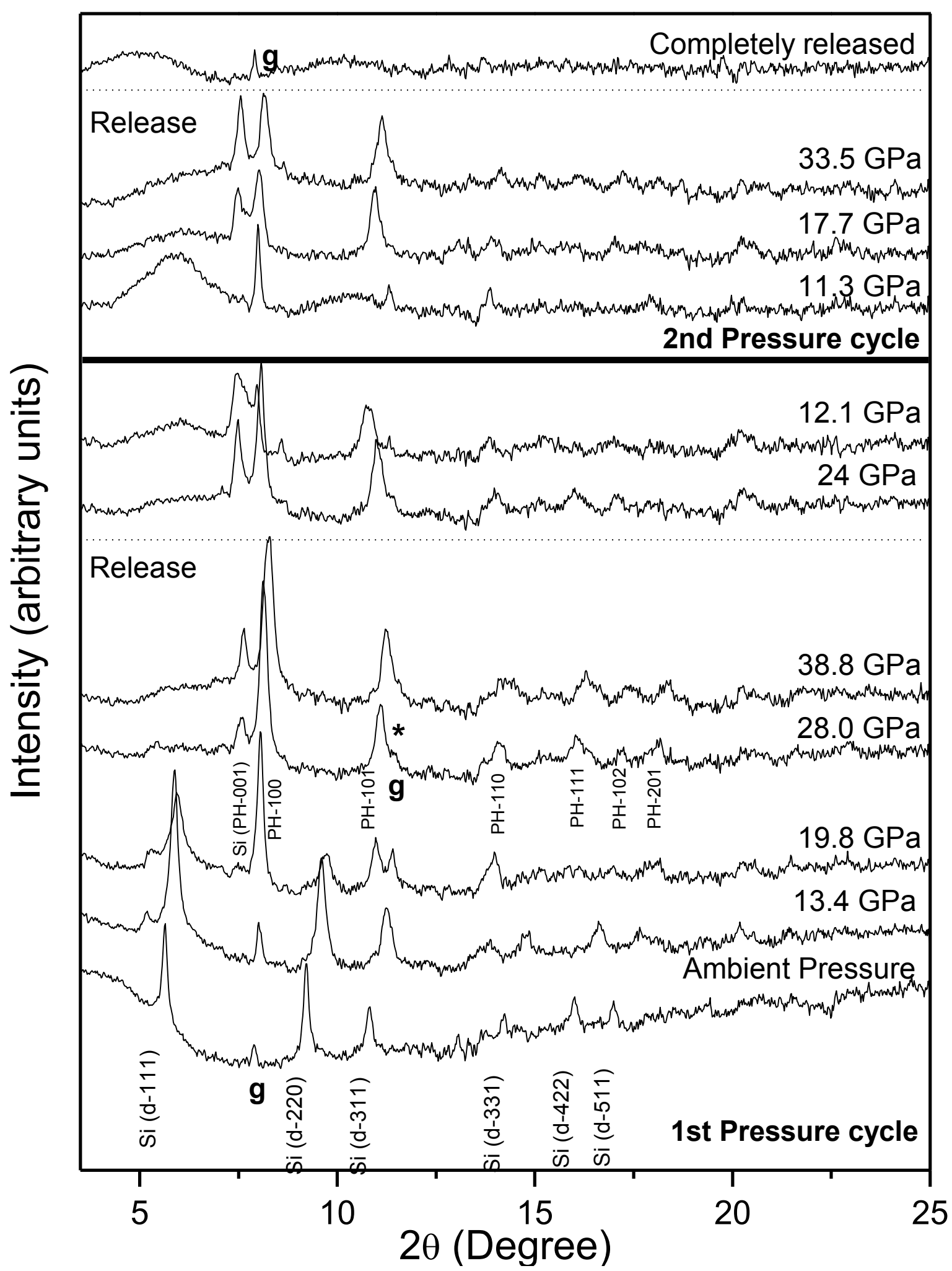

Fig. 2 The x-ray diffraction patterns of π-Si in the compression and decompression cycles. The background of the empty gasketed cell was subtracted from the diffraction patterns. The pattern at ambient pressure shows all the diffraction peaks of the cubic diamond phase of silicon. The tungsten gasket peaks have been marked with a * and have also been labeled as 'g'. The diffraction peaks of the high pressure phase have been indexed with the 'ph' structure.

## Figure 3

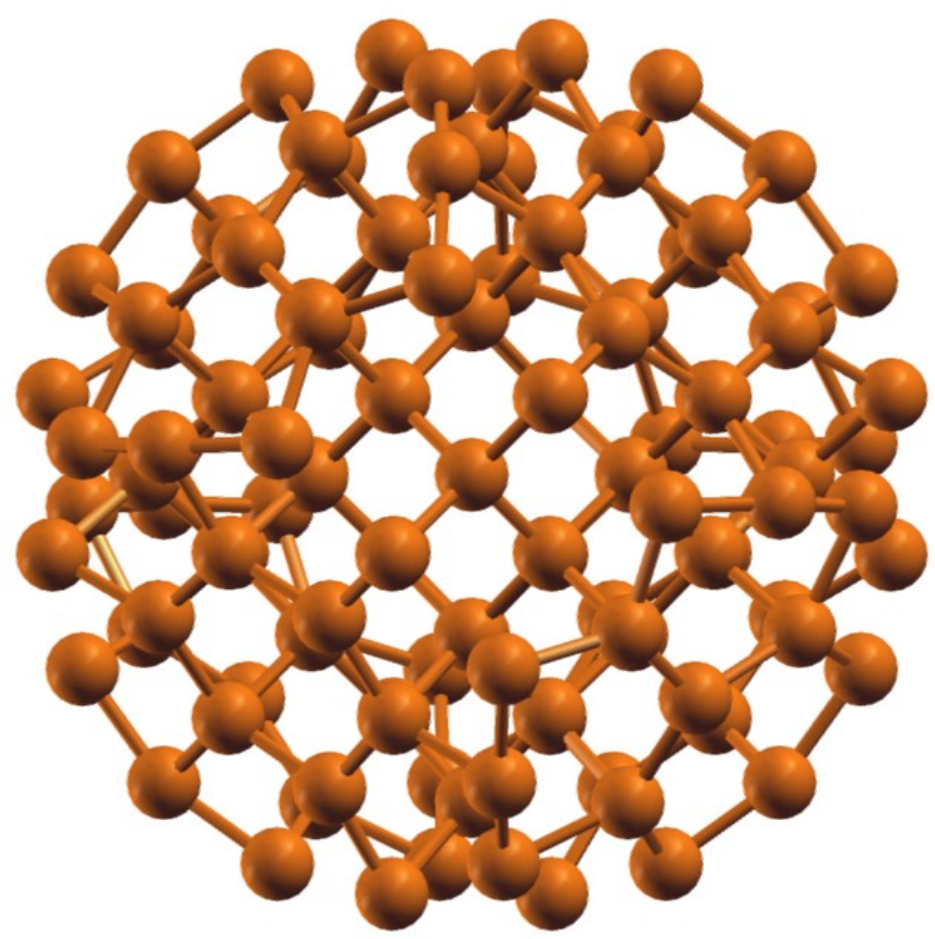

(a)

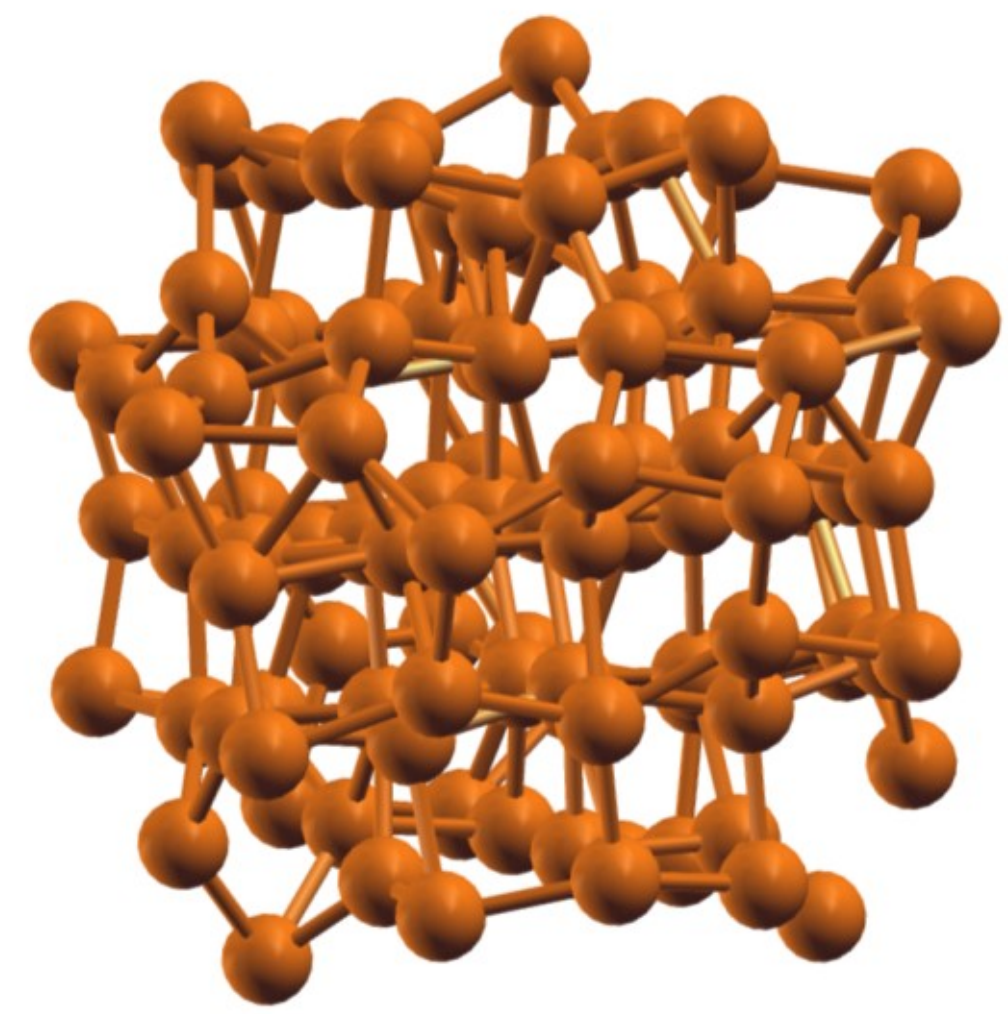

(b)

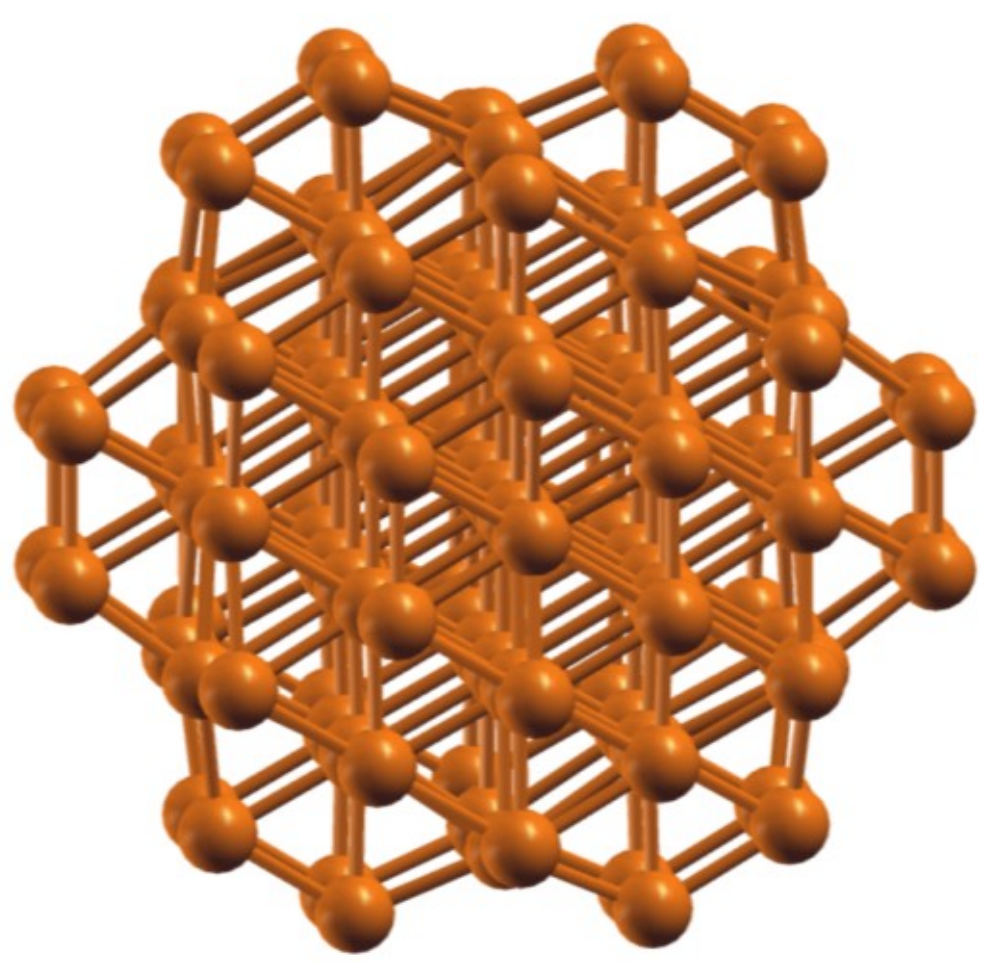

(c)

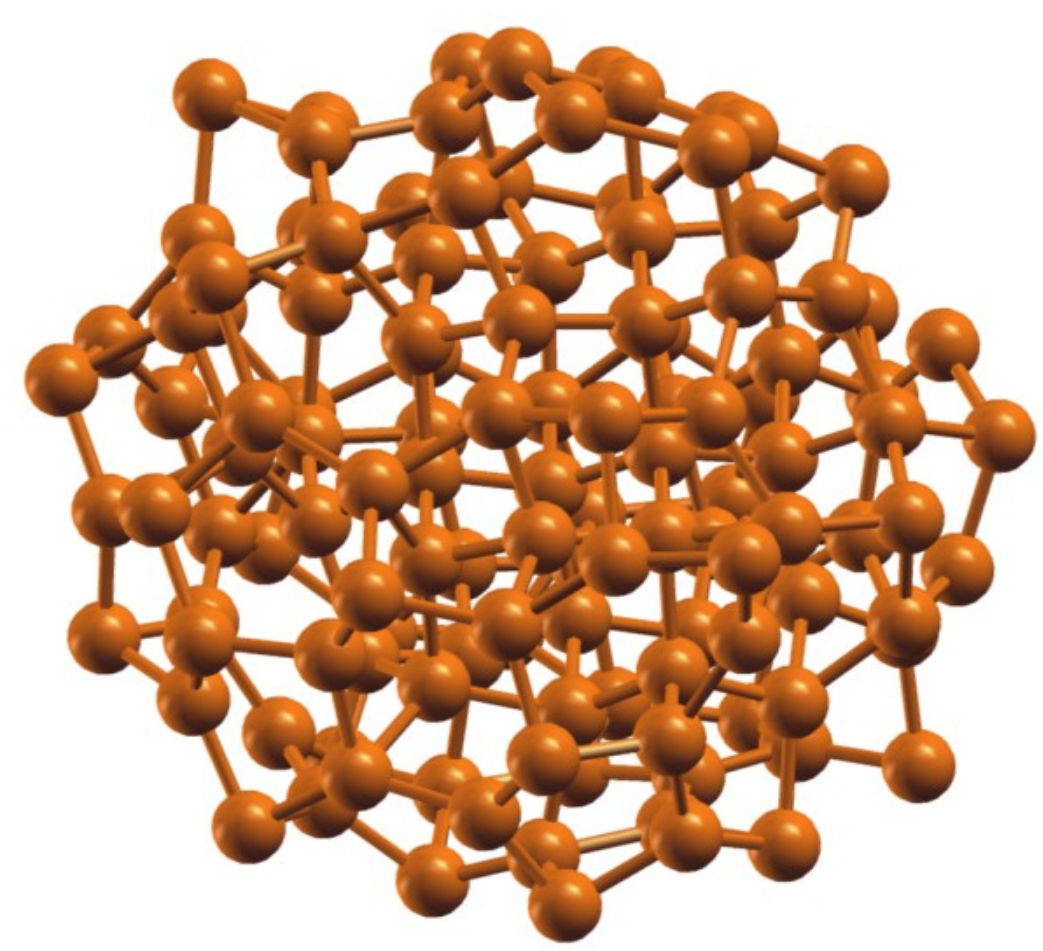

(d)

Fig. 3 Clusters of different structures of silicon as simulated by the first principles calculations. a) The cubic silicon cluster has the lowest energy among the stable crystal clusters. b) β-tin becomes disordered upon relaxation but retains somewhat higher coordination c) primitive hexagonal phase forms a more compact cluster although with a higher energy and d) the amorphous structure has almost the same volume as the diamond structure but has the lowest energy.